\documentclass[aps,preprint,superscriptaddress]{revtex4}
\usepackage{amsmath}
\usepackage{amsfonts}
\usepackage{amssymb}
\usepackage{graphicx}
\usepackage{epstopdf}
\usepackage{multirow}
\usepackage{tabularx}
\usepackage{ragged2e}

\newcommand{\BM}[1]{\mbox{\boldmath$#1$}}

\newcommand{\Bq}{{\BM q}}
\newcommand{\BI}{{\BM I}}

\newcommand{\Bsigma}{{\BM \Sigma}}

\newcommand{\Bepsilon}{{\BM \epsilon}}

\begin{document}

\title{Universal avalanche exponents in plasticity of amorphous materials}

\author{Zoe Budrikis}
\affiliation{ISI Foundation, Via Alassio 11/c, 10126 Torino, Italy}
\author{David Fernandez Castellanos}
\affiliation{WW8-Materials Simulation, FAU University of Erlangen-Nuremberg, Germany}
\author{S. Sandfeld}
\affiliation{WW8-Materials Simulation, FAU University of Erlangen-Nuremberg, Germany}
\author{M. Zaiser}
\affiliation{WW8-Materials Simulation, FAU University of Erlangen-Nuremberg, Germany}
\author{Stefano Zapperi}
\affiliation{Center for Complexity and Biosystems, Department of Physics, University of Milano, via Celoria 26, 20133 Milano, Italy}
\affiliation{CNR-IENI, Via R. Cozzi 53, 20125 Milano, Italy}
\affiliation{ISI Foundation, Via Alassio 11/c, 10126 Torino, Italy}
\affiliation{Department of Applied Physics, Aalto University, P.O.Box 11100
FI-00076 AALTO, Finland}

\begin{abstract}
{Plastic yield of amorphous solids occurs by power law distributed slip avalanches whose universality is still debated. 
Determination of the power law exponents from experiments and molecular dynamics simulations is hampered by limited statistical sampling. On the other hand, while existing elasto-plastic depinning models  give precise exponent values, these models to date have been limited to a scalar approximation of plasticity which is difficult to reconcile with the statistical isotropy of amorphous materials. Here we introduce for the first time a fully tensorial mesoscale model for the elasto-plasticity of disordered media that can not only reproduce a wide variety of shear band patterns observed experimentally for different deformation modes, but also captures the avalanche dynamics of plastic flow in disordered materials. Slip avalanches are characterized by universal distributions which are quantitatively different from mean field predictions, both regarding the exponents and regarding the form of the scaling functions, and which are independent of system dimensionality (2D vs 3D), boundary and loading conditions, and uni-or biaxiality of the stress state. We also measure average avalanche shapes, which are equally universal and inconsistent with mean field predictions. Our results provide strong evidence that the universality class of plastic yield in amorphous materials is distinct from that of mean field depinning.}
\end{abstract}

\maketitle

\section*{Introduction}
Plasticity of both crystalline and amorphous materials is interesting from the point of view of statistical mechanics of complex systems because, during irreversible deformation under load, solids exhibit a rich variety of collective phenomena including spontaneous strain localization, intermittent dynamics and power-law distributed avalanches. This phenomenology overlaps with that seen in systems as diverse as ferromagnets exhibiting Barkhausen noise~\cite{Durin2006} and crack propagation~\cite{Bonamy2008,Laurson2010,Laurson2013}, and a unified conceptual model for these problems is desirable. While there is consensus that plastic deformation in amorphous materials occurs via local reorganizations~\cite{Argon1979101, Argon1983499, Falk1998, Albaret2016}, how these behave collectively is a topic that continues to receive considerable experimental and theoretical attention~\cite{Baret2002,Picard2002,Picard2005, Jagla2007, Shan2008,Dahmen2009,Wang2009, Homer2010, Sun2010,Sun2011,Martens2011, Talamali2011,Talamali2012, Sun2012,Budrikis2013,Salerno2013, Nicolas2014, Lin2014a, Lin2014, Antonaglia2014, Antonaglia2014a,Liu2016}. 

A key point of debate concerns the universality class of the plastic yield transition. As the external drive reaches a critical value the system undergoes a non-equilibrium phase transition into a flowing phase. The essential physics of this transition is captured by depinning-like models in which plastic deformation with fluctuating local yield thresholds is mapped to the motion of an interface through a disordered medium \cite{Talamali2011,Budrikis2013,Lin2014,Lin2015,Sandfeld2015,Liu2016}. Mean-field depinning models provide predictions for critical exponents which have been tested in atomistic~\cite{Salerno2013,Liu2016} or mesoscale simulations \cite{Baret2002,Picard2002,Picard2005,Talamali2011,Budrikis2013,Lin2014,Lin2015,Liu2016,Sandfeld2015} and in experiments \cite{Wang2009,Sun2010,Antonaglia2014,Antonaglia2014a}, but results remain inconclusive. For example, reported values of the avalanche size exponent $\tau$ range from $1.25$ to $1.5$,
often with large error bars. Faced with such data, different groups have interpreted the results as either consistent with the mean-field exponent of $\tau=3/2$ \cite{Antonaglia2014,Antonaglia2014a} or not \cite{Talamali2011,Budrikis2013,Lin2014,Lin2015,Sandfeld2015}, and controversy continues.

Many molecular dynamics and experimental studies face limitations in terms of accessible system size and strain rate range. Often, studies are limited to a narrow range of avalanche sizes for such technical reasons, with scaling regimes spanning a single decade only, making accurate determination of exponents difficult if not impossible. Furthermore, as has been demonstrated recently~\cite{Liu2016}, it is essential to ensure applied strain rates are slow enough to be truly adiabatic in order to avoid exponent drift. 

While these problems can be overcome by lattice based mesoscale models, existing mesoscale models often treat the plastic strain as a scalar variable --- effectively assuming that shear occurs in a single direction only  \cite{Baret2002,Picard2002,Picard2005,Talamali2011,Martens2012,Budrikis2013,Nicolas2014,Lin2014,Lin2015,Sandfeld2015}. This assumption which is inherent in all scalar models sits very uneasily with the statistically isotropic nature of amorphous solids and makes such models appear more suited to single-slip crystal plasticity \cite{Zaiser2005}. In case of finite samples subject to general loads, the assumption of a scalar strain occurring in response to a scalar load is quite obviously at variance with the tensor nature of stress and strain in continuum mechanics. This assumption makes it in principle impossible to model general deformation processes where stress is heterogeneous and multi-axial. Even if the applied stress is acting along a single stress axis, the models predict spontaneous strain localization and spatially fluctuating plastic strain fields, which implies that to ensure compatibility of deformation the stress field must locally be always multi-axial (see e.g. \cite{Sandfeld2014}). So one may legitimately ask what is the relevance of predictions derived from scalar models to real-world plasticity. 

Tensorial plasticity models have been regularly used in the context of amorphous materials (see e.g. \cite{Bulatov1994,Anand2005,Su2006,Yang2006,Homer2009}). However, these models -- even where they are parameterized based upon microscale simulations \cite{Schuh2003,Homer2009} and consider statistical flow rules \cite{Bulatov1994,Homer2009} -- tend to assume spatially homogeneous constitutive equations which do not fully reflect the stochastic heterogeneity of plastic deformation properties on the smallest scales. As a consequence, while such models are able to predict complex deformation states on the macroscale, they are not capable of adequately capturing the avalanche dynamics of plastic deformation. We are thus facing a conundrum: On the one hand we have (over-)simplified statistical physics models of avalanche dynamics which because of their scalar nature cannot capture real-world deformation processes, on the other hand we have continuum mechanics models which because of their homogeneous and/or deterministic nature cannot capture avalanche dynamics of plastic flow. As avalanche phenomena are assumed to play a key role in the early stage of shear band formation, which in turn controls the macroscopic deformation behavior of many amorphous materials, this deficiency may prevent a comprehensive theoretical understanding of amorphous plasticity. 

In the present paper we address this fundamental issue by formulating a fully tensorial model of plasticity of disordered solids which accounts for the discrete and stochastic nature of the elementary deformation processes on the smallest scales and thus bridges the gap between statistical physics and engineering plasticity approaches.

We apply the model to both two- and three-dimensional problems, including finite samples, intrinsically heterogeneous deformation processes such as bending and indentation, and genuinely multi-axial loading conditions.  For all these systems we analyze the avalanche statistics in terms of the stress dependent distribution of avalanche sizes and determine the exponent $\tau$ which controls the power-law regime of the avalanche size distributions, the exponent $\sigma$ which controls the divergence of the maximum avalanche size near the yield stress, and the exponent $\gamma$ which connects duration and size of avalanches.  Our simulations yield the surprising result that, under all these circumstances, the exponents $\tau$ and $\gamma$ neither depend on the nature of the stress and strain variables (scalar/uniaxial vs. tensorial/multiaxial), nor on the dimensionality of the system (2D vs. 3D), nor on the presence or absence of macroscopic strain gradients. Where $1/\sigma$ can be meaningfully measured, it is also the same for all loading conditions. Furthermore, our avalanche distributions can be fitted quantitatively by non-mean field scaling functions. Finally, we also demonstrate that not only are avalanche size distributions universal and not mean field, but so are the average shapes of avalanche signals.

\section*{Model}
Our plasticity model is formulated in the spirit of rate-independent continuum plasticity with a J2/Von Mises type yield criterion, which we generalize to account for structural randomness and deformation occurring in localized, discrete events. We implement our simulations on a $D$ dimensional cubic lattice and assign to each lattice element a local stress tensor $\Bsigma$. From this we define a yield condition as $\Sigma_{\rm eq} = \sqrt{(3/2) \Bsigma':\Bsigma'} > \Sigma_y$,
where $\Bsigma' = \Bsigma - (1/D) {\rm Tr} \Bsigma \BI$ is the deviatoric part of the stress tensor and $\BI$ is the rank-2-unit tensor in $D$ dimensions. Activation of a shear event occurs as soon as the equivalent stress at a lattice site exceeds the randomly assigned threshold $\Sigma_y$ which we draw from a uniform distribution over $[0,1)$. Such activation results in local atomic re-arrangements which lead to a change of the local yield threshold, which is newly assigned after each event. In this respect our model differs significantly from the zero-temperature limit of stochastic plasticity models which consider thermal activation of shear events with a stress dependent threshold, as assumed in the Kinetic Monte Carlo approaches proposed by Bulatov and Argon \cite{Bulatov1994} and Homer and Schuh \cite{Homer2009}. These models assume a fixed and uniform local yield threshold, an assumption which may not fully capture the influence of atomic-scale randomness on plastic flow and makes such models unsuitable for investigating avalanche phenomena. 

The stress on an element is the sum of internal stress arising from the plastic strain field $\Bepsilon^{\rm p}$ and an applied load. In simulations with periodic boundary conditions (PBCs), the loading is a spatially homogeneous `external' stress field $\Bsigma^{\rm ext}$ which is understood to arise from remote boundary tractions applied to the infinite contour. Simulations of finite samples are performed using standard finite element methods (FEM) to solve the elastic-plastic problem with the specified boundary displacements and/or tractions. For details of the FEM implementation in the context of a scalar model, see \cite{Sandfeld2015}. In both cases, the external stress is increased adiabatically slowly and is held constant during avalanches, as described previously~\cite{Budrikis2013, Sandfeld2015}. When an element yields, we increase its local plastic strain $\Bepsilon^{\rm p}$ by an amount $\Delta \Bepsilon^{\rm p} = \hat{\Bepsilon}\Delta \epsilon $ where the strain direction $\hat{\Bepsilon} = \Bsigma'/\Sigma_{\rm eq}$ is chosen to maximize the locally dissipated energy and $\Delta \epsilon$ is a fixed strain increment. The size $S$ of an avalanche is defined as the to tal number of local strain increments, and its duration $T$ is the number of parallel update steps performed. 

The `internal' stress arising from the plastic strain field is calculated using either a Green's function method (for PBCs) or FEM (for finite systems). In the former case we consider stress and strain only at the element center-points which form a cubic grid with periodic boundary conditions. In Fourier space, the internal stress field $\Bsigma^{\mathrm{int}}$ is given by $\Bsigma^{\mathrm{int}}_{ij}(\Bq) = \boldsymbol{G}_{ijkl}(\Bq) \Bepsilon_{kl}^{\rm p}(\Bq)$.
The interaction kernel $\boldsymbol{G}$ is obtained by treating the plastic strain of an element as the strain of an Eshelby inclusion of vanishing volume located at the element centre-point, for which the stress field is known analytically~\cite{Eshelby1957}. Continuing this solution periodically with period $L$ allows us to use a Fourier transform to obtain the overall internal stress field that arises from superposition of all element stresses.

For finite samples, our FEM implementation uses four-node linear elements of square shape. Each element is associated with an element stress that is evaluated as the average stress over the element. An active element experiences a plastic strain increment that is homogeneous over the element and zero elsewhere. The models are matched by ensuring that the plastic strain field of the point-like Eshelby inclusion, integrated over the element, has the same value as the homogeneous element strain in the finite element model.

Simulations have been performed in both two and three dimensions, considering both macro-homogeneous deformation modes (uniaxial tension with free surfaces and applied tensile tractions, pure shear, bi-axial deformation with applied tensile and shear tractions) and, in 2D, also macro-heterogeneous deformation modes such as simple shear, bending, and indentation. Systems in 3D had $32^3$ elements, and 2D systems had $256^2$ elements.

\section*{Results}
\subsection*{Depinning models give strain localization dependent on loading conditions.}
\begin{figure*}
\includegraphics[width=0.6\columnwidth]{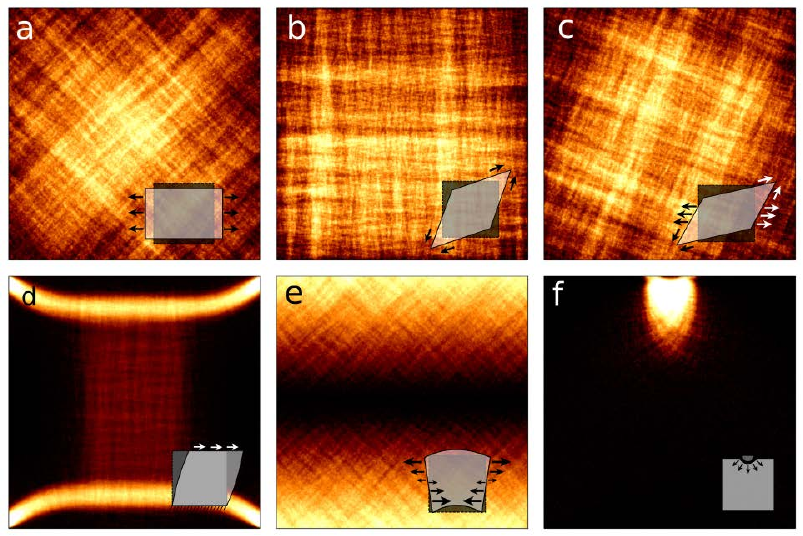}
\includegraphics[width=0.59\columnwidth]{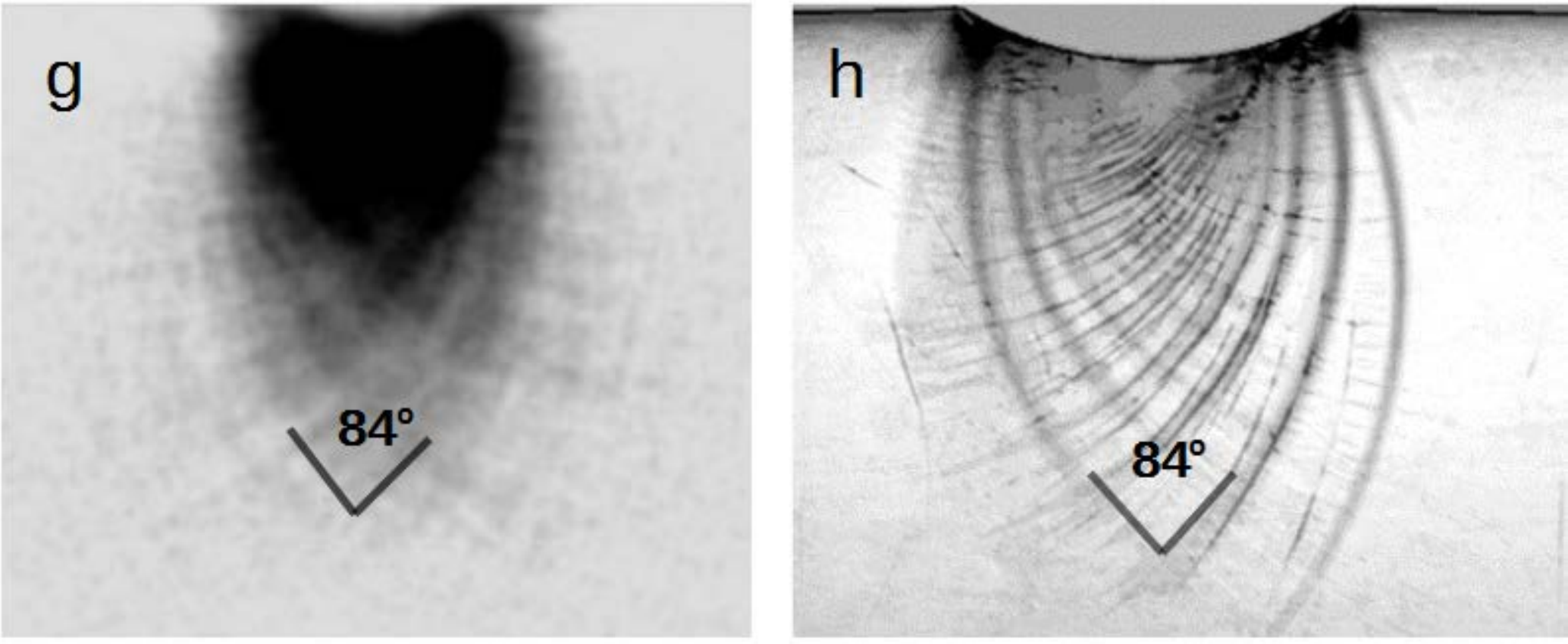}
\caption{Strain localization patterns depend on deformation mode and are consistent with experimental studies. Shown here are typical configurations from 2D systems of size $256\times256$ with free surfaces. The loading conditions are a) pure tension; b) pure shear; c) biaxial loading; d) simple shear; e) bending; f) indentation; g) shows a magnification of figure (f) indicating the shear band intersection angle, g) experimental image of a shear band pattern in plane strain indentation, after \cite{Su2006}.}
\label{fig:patterns}
\end{figure*}

We first verify that our model gives rise to strain localization consistent with previous experimental and molecular dynamics studies~\cite{Vaidyanathan2001,Ramamurty2005,Shi2007,Maloney2009,Chen2010}. As illustrated in Fig.~\ref{fig:patterns}, the plastic strain field organizes into localization patterns (shear bands) which approximately follow the directions of maximum shear stress. In the case of uniaxial tension and biaxial loading, this is at approximately $45^{\circ}$ to the tensile axis. In the case of simple shear the vertically fixed surfaces cause strong stress concentrations in the specimen corners. Under simulated 2D indentation with a circular indenter, strain localizes into a pattern of intersecting circles. This patterrn is typical of shear bands observed in indentation of bulk metallic glasses \cite{Su2006}, and the simulations correctly reproduce even details of the incipient shear band pattern such as the slightly acute intersection angle of the shear bands (Fig.~\ref{fig:patterns}g,h). We emphasize that this type of strain patterning cannot be captured at all by scalar plasticity  models.

\subsection*{Avalanche exponents are universal for a variety of loading and boundary conditions.}
We now turn to measurement of the exponents describing the avalanche size distributions. While in mean field depinning, the avalanche size distribution is given by $P(S)\sim S^{-\tau} \exp(-S/S_0)$, we find that this simple exponential tail does not fit our data. We turn instead to a first-order correction to the mean field size distribution, given by~\cite{LeDoussal2012}
\begin{equation}
\label{eq:avalancheDistribution}
P(S) = \frac{A}{2\sqrt{\pi}} S^{-\tau} \exp\bigl(C\sqrt{u} - \frac{B}{4} u^{\delta}\bigr).
\end{equation}
Here, $u = S/S_{\rm max}$, where the size $S_{\rm max}$ of the largest avalanches diverges like $S_{\rm max} \propto (\Sigma_{\rm c} - \Sigma)^{-1/\sigma}$ as the external loading $\Sigma$ (in simulations without periodic boundary conditions, averaged over the loaded face) approaches a non-universal critical value $\Sigma_c$, and $1/\sigma$ is a universal exponent. The parameters $A$, $B$, $C$ and $\delta$ are, in terms of the exponent $\tau$, given by~\cite{LeDoussal2012}
\begin{subequations}
\begin{gather}
A = 1 + (2-3\gamma_E)(\tau-3/2)/3,\\
B = 5 + \gamma_E - 2\tau(4-\gamma_E)/3,\\
C = 2\sqrt{\pi} - 4\sqrt{\pi}\tau/3,\\
\delta = 2(1-\tau/3),
\end{gather}
\end{subequations}
where $\gamma_E\approx0.577216$ is Euler's constant.

\begin{figure*}[hbp]
\includegraphics[width=0.8\columnwidth]{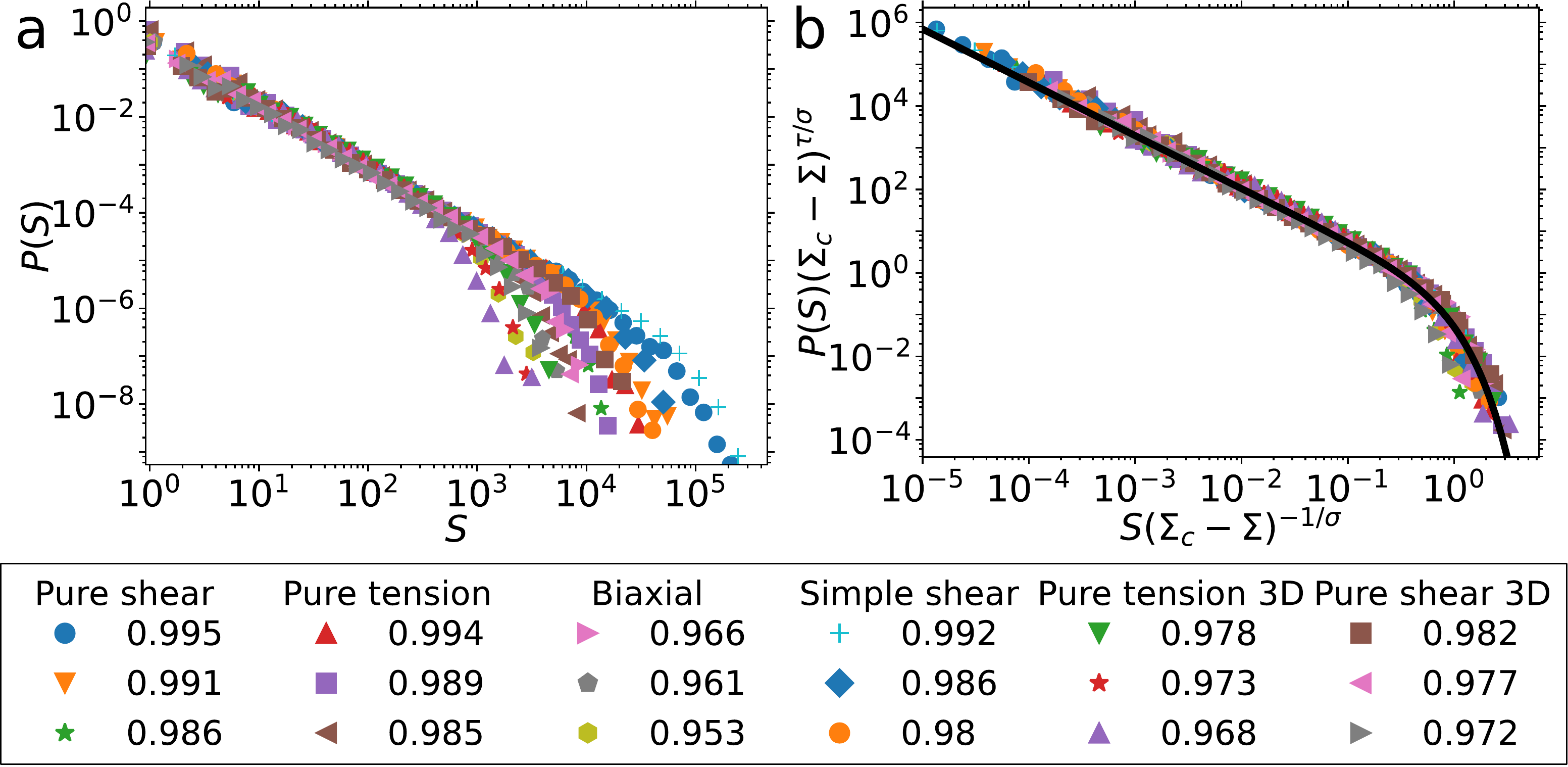}
\caption{Avalanche size distributions are universal under changes in dimensionality and loading, with exponents $\tau=1.28\pm0.003$ and $1/\sigma=1.95\pm0.01$. Panel (a) shows avalanche size distributions for all loading conditions at different external stresses, panel (b) shows the same distributions collapsed using the exponents $\tau$ and $1/\sigma$, which have been measured by a joint fit of all data sets. The black line is the theoretical distribution of Ref.~\cite{LeDoussal2012}, given in Eq.~\eqref{eq:avalancheDistribution}. The values in the legend refer to the ratio $\Sigma/\Sigma_C$ for each loading condition, the fitted $\Sigma_c$ values are given in Table~\ref{fc}.}
\label{fig:avalstats}
\end{figure*}

We use the form \eqref{eq:avalancheDistribution} to simultaneously fit avalanche size distributions for six different loading/boundary conditions: (i) Uniaxial tension in 3D with periodic boundary conditions; (ii) uniaxial tension in 2D with free side surfaces; (iii) biaxial deformation in 2D with superimposed tensile and shear tractions; (iv) pure shear in 3D with periodic boundary conditions; (v) pure shear in 2D with uniform shear tractions applied to the unconstrained side surfaces; (vi) simple shear in 2D with horizontal traction forces applied to the vertically constrained top surface and with fixed bottom surface. We emphasize that the only fitting parameters are $\Sigma_c$ for each loading condition, and the exponents $\tau$ and $1/\sigma$, which are the same for all loading conditions, with no additional free parameters. We find $\tau=1.28\pm0.003$ and $1/\sigma=1.95\pm0.01$. Fig.~\ref{fig:avalstats} shows the measured avalanche size distributions and their collapse using these two parameters, along with the fit of \eqref{eq:avalancheDistribution} to the collapsed data. Table~\ref{fc} gives the fitted $\Sigma_c$ values for each loading.

To test whether a joint fit of avalanche distributions for different loading conditions is appropriate, we have also fitted each loading condition separately with the functional form \eqref{eq:avalancheDistribution}. Data collapses from these fits are shown in Fig.~S1. The fitted exponent $\tau$ is nearly identical across the loading conditions, with separate fits giving $\langle\tau\rangle=1.26\pm0.01$ and values ranging from $1.25$ to $1.28$. 

On the other hand, when fitted separately for each loading condition, $1/\sigma$ varies considerably, from $1.53$ for biaxial loading to $2.05$ for pure tension in 2d, with mean fitted value $\langle1/\sigma\rangle=1.8\pm0.2$. This certainly indicates that the nominal errorbar of $0.01$ in the joint fit is an underestimate. However, because of the strong universality of the other exponents as well as the avalanche shapes (see below), a joint fit of all loading conditions is an appropriate procedure. Indeed, since $1/\sigma$ and $\Sigma_c$ both determine $S_{\mathrm{max}}$ for each distribution and both are fitted, a range of $1/\sigma$ and $\Sigma_c$ values will give fits of approximately equal quality, and constraining $1/\sigma$ to be the same for all loadings avoids spurious over-fitting of individual distributions. This interpretation is also supported by the high quality of the data collapse in Fig.~\ref{fig:avalstats}(b).

Using our finite-element based code, we have also studied more complex loading conditions such as bending and indentation. Bending corresponds to a heterogeneous but uniaxial stress state; however, hardening is observed (Fig.~S2). For indentation, the stress state is both heterogeneous and multi-axial and in addition the fraction of the simulated specimen that actually undergoes plastic deformation increases with increasing force acting on the indenter. Accordingly, in both cases, no critical stress $\Sigma_c$ can be defined. 

\begin{figure}
\includegraphics[width=0.7\columnwidth]{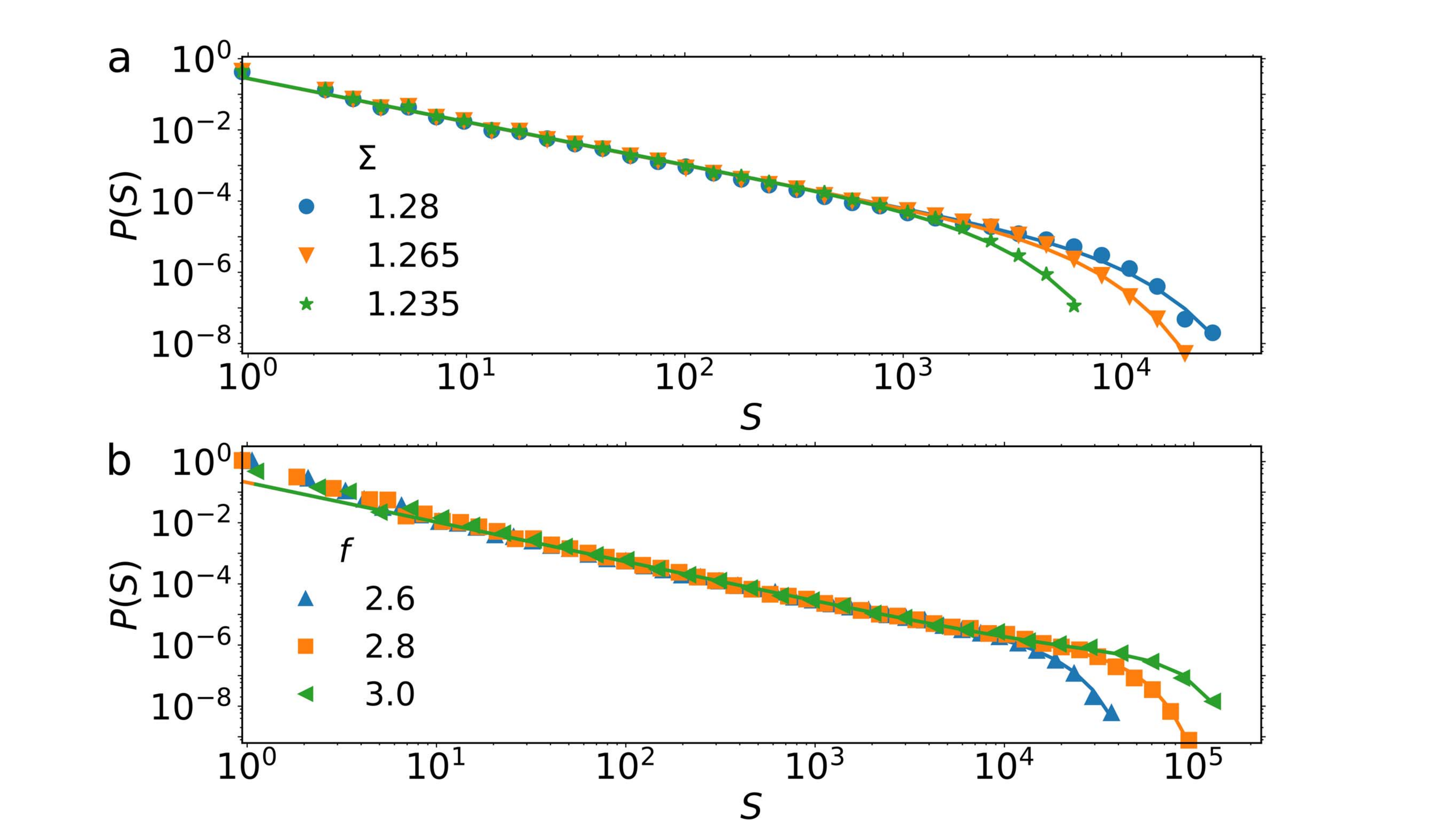}
\caption{Avalanche size distributions for bending and indentation are governed by the same exponent $\tau\approx1.28$ as for other loading conditions. Panel (a) shows avalanche size distributions for bending. Lines are fits using the form given in Eq.~\eqref{eq:avalancheDistribution}, but fitting $S_{\mathrm{max}}$ independently for each $\Sigma$, to yield $\tau=1.221\pm0.004$. Panel (b) shows avalanche size distributions for indentation with a spherical indenter. Here $f$ denotes the force acting on the indenter divided by the indenter cross section. Lines are fits using the form given in Eq.~\eqref{eq:indentfits}, and we measure $\tau=1.29\pm0.01$.}\label{fig:indentation}
\end{figure}

However, despite these complexities, the avalanche distribution is described by an exponent $\tau$ consistent with other cases, as shown in Fig.~\ref{fig:indentation}. We have fitted the distributions to obtain $\tau$. In the case of bending, the fitting form of Eq.~\eqref{eq:avalancheDistribution} is used, with $\tau$ shared between all curves but with $S_{\rm max}$ fitted independently for each $\Sigma$. We find $\tau=1.221\pm0.004$. For indentation, the distributions display distinct `bumps' in the tails and we use a fitting form
\begin{equation}
\label{eq:indentfits}
P(S)=a S^{-\tau} \exp(-bS^2+cS),
\end{equation}
with $a$ and $\tau$ shared between all curves, and $b$ and $c$ fitted independently for each $\Sigma$. Here, we find $\tau=1.29\pm0.01$.

\begin{table}[]
\centering
\begin{tabular}{|l|c|}
\hline
Loading & $\Sigma_c$ \\ \hline
Pure shear, 2D & $0.57746 \pm 7\times10^{-5}$ \\
Pure tension, 2D & $1.13500 \pm 9\times10^{-5}$\\
Biaxial, 2D & $0.4124 \pm 3\times 10^{-4}$\\
Simple shear, 2D & $0.31591 \pm 8\times10^{-5}$\\
Pure tension, 3D, PBCs & $0.9390 \pm 3\times 10^{-4}$\\
Pure shear, 3D, PBCs & $0.4944 \pm 2\times 10^{-4}$\\
\hline
\end{tabular}
\caption{Nonuniversal fitted yield stress $\Sigma_c$ for the loading conditions shown in Fig.~\ref{fig:avalstats}, fitted using Eq.~\eqref{eq:avalancheDistribution}.}
\label{fc}
\end{table}

Remarkably, although the strain patterns depend strongly on the loading conditions, the avalanche exponent $\tau$ is not only independent of the dimensionality (2D vs. 3D),  but it is also independent of whether the stress field is homogeneous and uniaxial (pure shear, pure tension), homogeneous and biaxial, or inhomogeneous and uni- or biaxial (bending, simple shear, indentation). Pure tension and pure shear produce identical results, which indicates that the avalanche exponent is not influenced by the presence or absence of hydrostatic stresses (pure tension adds a hydrostatic stress contribution), by boundary conditions, or by the orientation of the simulation grid (deviatoric stresses are in both cases equivalent but the directions of shear differ by 45 degrees).

\begin{figure}
\includegraphics[width=20pc]{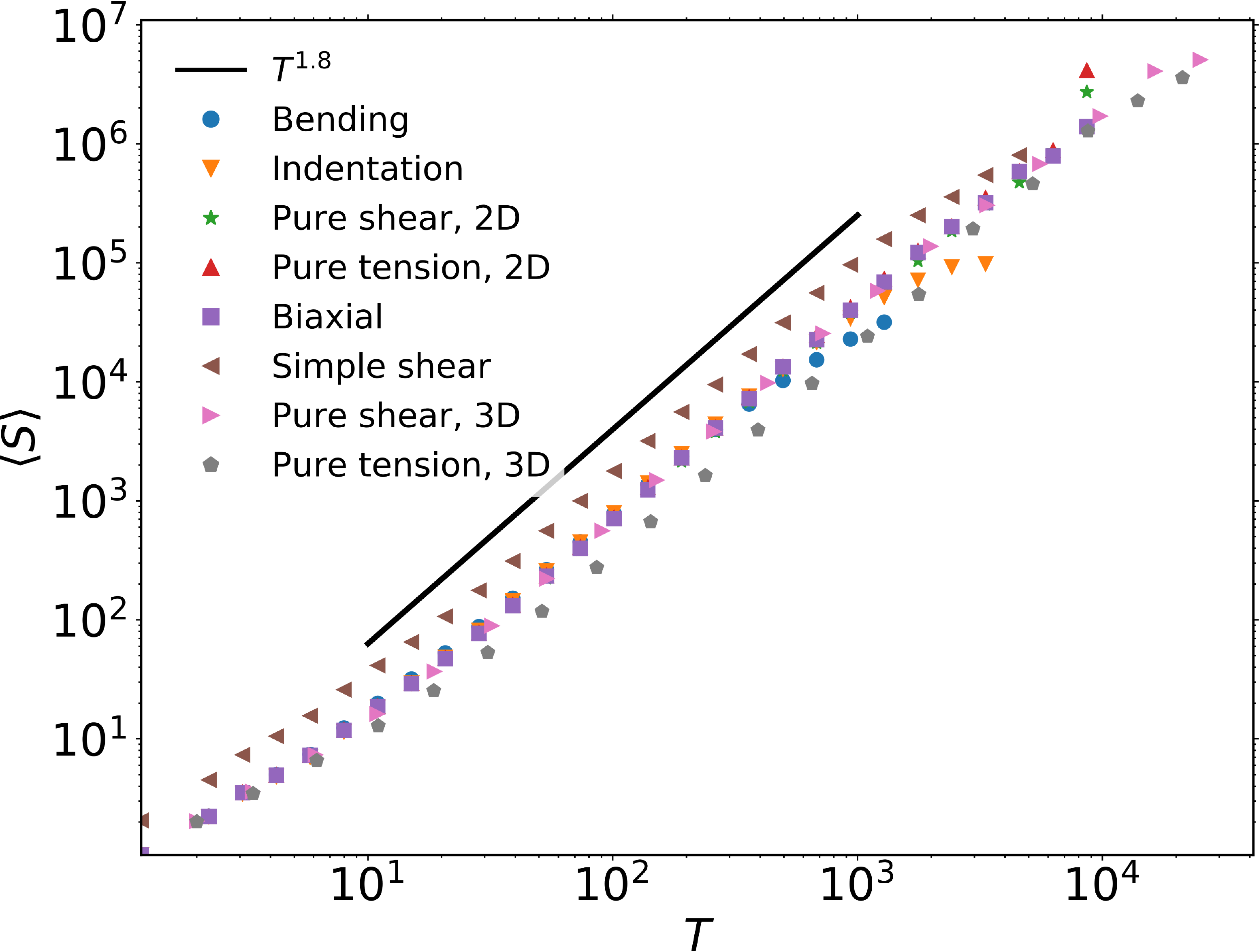}
\caption{Avalanche size $\langle S \rangle$ {\it vs} duration $T$ is governed by a power law $\langle S \rangle \propto T^{\gamma}$, $\gamma=1.8\pm0.1$, for different deformation conditions and system dimensionalities. \label{fig:timesize}}
\end{figure}

A third universal exponent $\gamma$ defines the relationship between avalanche duration $T$ and avalanche size $S \propto T^{\gamma}$. To determine this exponent we define the duration of an avalanche as the number of simultaneous updates required from the start of the avalanche to the moment when all elements are stable again. Also here we find strong universality with an exponent $\gamma = 1.8 \pm 0.01$ for all loading conditions, including bending and indentation (Fig. \ref{fig:timesize}).

\subsection*{Avalanche temporal signals are not consistent with mean field theory.}
Beyond avalanche size distributions, the average temporal signal of avalanches, $\langle \dot{\epsilon}\rangle$, is also expected to take a form that depends on the universality class of the yield transition~\cite{Laurson2013, Dobrinevski2014}. 

We first measure the average avalanche shape for avalanches of fixed duration $T$. To first order, for time $t'=t/T$ and with normalization such that the area under the curve is $1$, the expected form is~\cite{Laurson2013}
\begin{equation}
\label{eq:avshape}
\langle \dot{\epsilon}(t') \rangle = (t'(1-t'))^{1-\gamma}(1-a(t'-1/2))/B(\gamma,\gamma),
\end{equation}
where $\gamma$ is the avalanche size {\it vs} duration exponent measured above, $B(x,y)=(x-1)!(y-1)!/(x+y-1)!$ is the beta function, and $a$ is a parameter describing the asymmetry of the avalanche shape. 

\begin{figure}
\includegraphics[width=0.5\columnwidth]{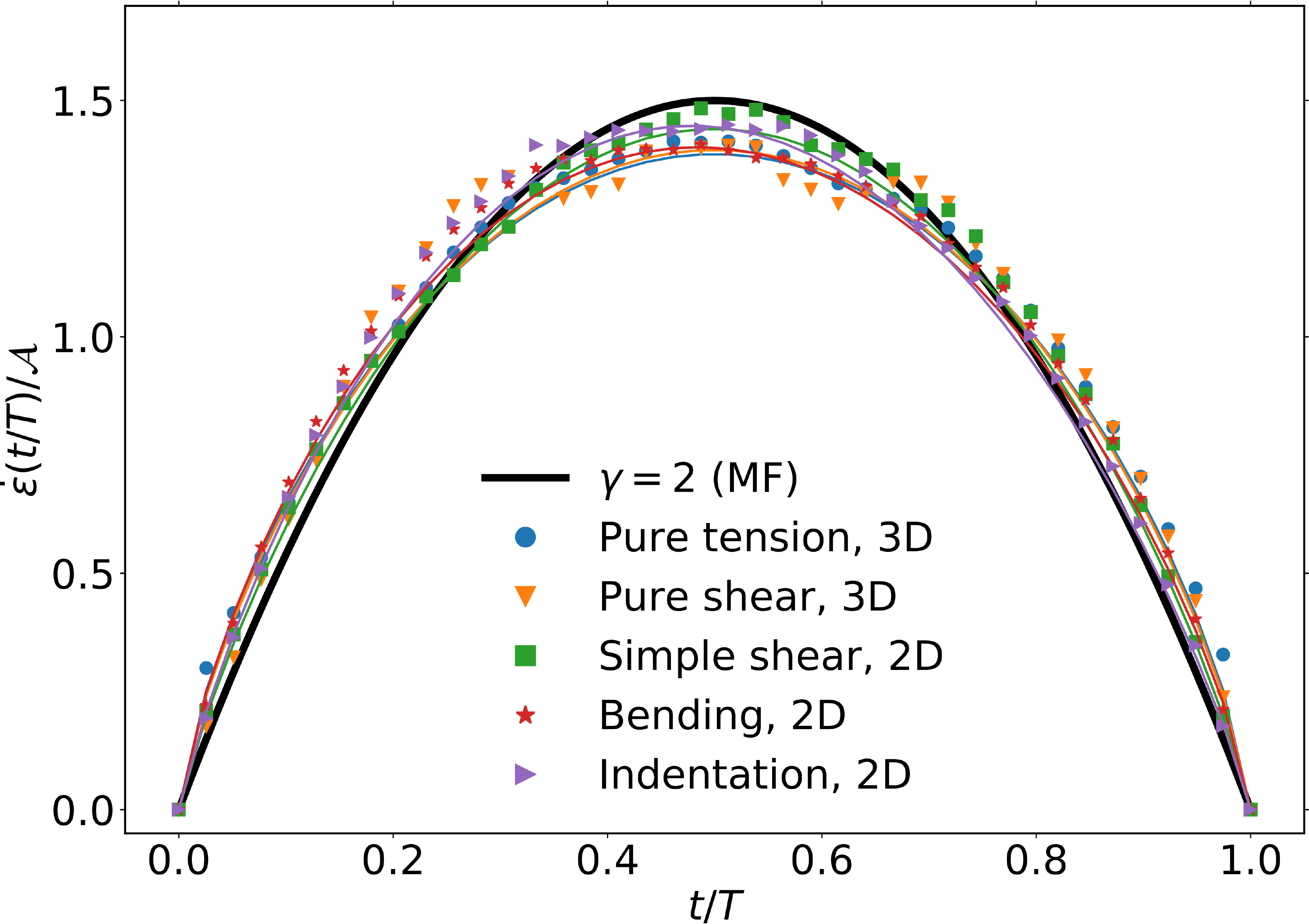}
\caption{The average avalanche signal for fixed duration $T$ is consistent with $\gamma=1.8$ and inconsistent with mean field theory. The signal has been averaged over avalanche realizations for durations $T=40,100,105,110,\ldots,200, 250$, and the mean signal for each $T$ is normalized so the area under the curve $\dot{\epsilon}(t/T)$ is unity before averaging over durations. Lines are fits using \eqref{eq:avshape}; fitted parameters are reported in Table~\ref{tab:meanshape}. \label{fig:meanshape}}
\end{figure}

We have measured the average shape of avalanches of fixed duration under several loading conditions: pure tension and pure shear in systems with periodic boundary conditions in 3D, as well as simple shear, bending and indentation in systems with free surface boundary conditions in 2D. As shown in Fig.~\ref{fig:meanshape}, we fit each curve using \eqref{eq:avshape} to obtain a mean $\gamma=1.8\pm0.05$, identical to the measurement based on avalanche size {\it vs} duration reported in Fig.~\ref{fig:timesize}, and inconsistent with the inverted parabola of mean field theory where $\gamma=2$ and $a=0$. A full list of fit parameters is given in Table~\ref{tab:meanshape}.

\begin{table}[]
\centering
\begin{tabular}{|l|c|c|}
\hline
Loading & $\gamma$ & $a$ \\ \hline
Pure tension, 3D & $1.74\pm0.02$ & --  \\
Pure shear, 3D & $1.74\pm0.02$ & --  \\ 
Simple shear, 2D & $1.86\pm0.02$ & --  \\ 
Bending, 2D & $1.77\pm0.02$ & $0.11\pm0.03$  \\ 
Indentation, 2D & $1.87\pm0.02$ & $0.15\pm0.03$  \\ 
\hline
\end{tabular}
\caption{Fitted exponent $\gamma$ and asymmetry parameter $a$ for the curves shown in Fig.~\ref{fig:meanshape}. Where $a$ values have been left blank, the asymmetry is too small to be measured by fitting the curves.}
\label{tab:meanshape}
\end{table}

\begin{figure}
\includegraphics[width=0.7\columnwidth]{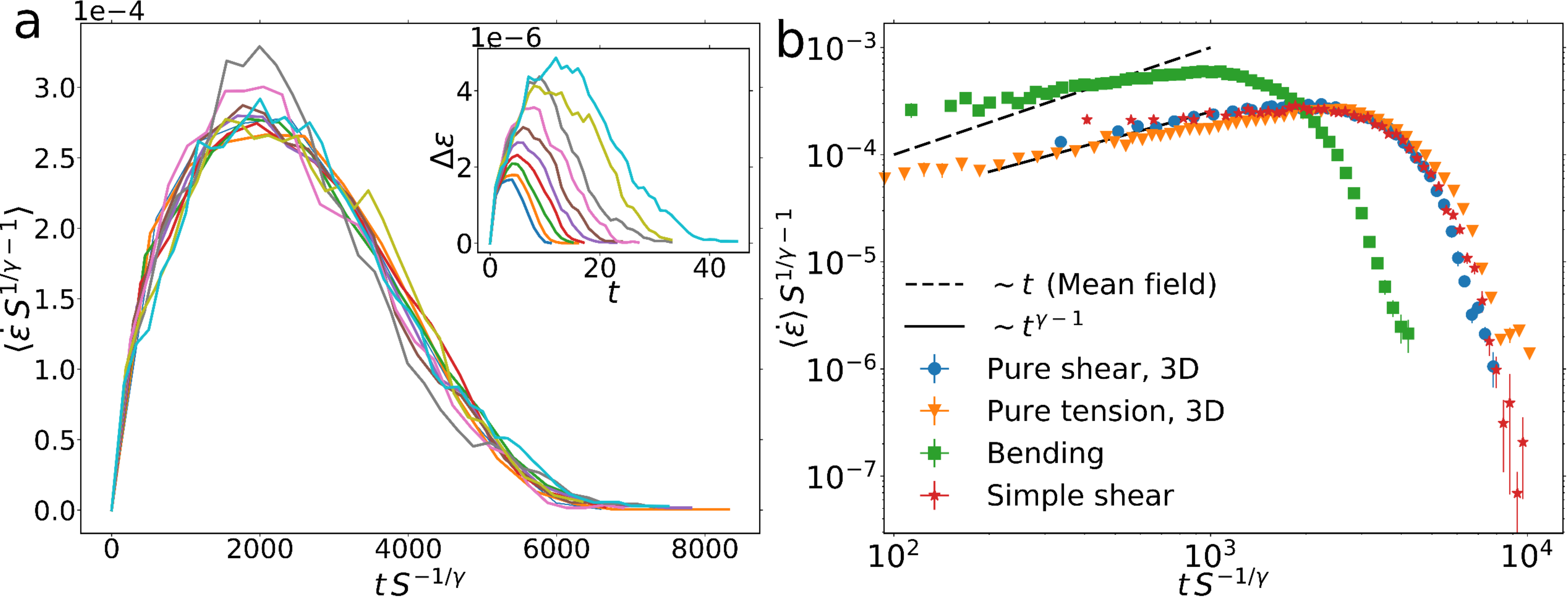}
\caption{Avalanche shapes for fixed size $S$ are inconsistent with the predictions of mean field theory. Panel a shows an collapse of avalanches of size $S$, $\log_{10}S=-5,-4.9,-4.8,\ldots,-4$, according to the scaling form \eqref{eq:avshapefixedS} with $\gamma=1.8$ as measured from avalanche size {\it vs} duration data reported in Fig.~\ref{fig:timesize}. The inset shows the original data. Panel b shows the averaged collapsed shapes for four different loading/boundary conditions. Contrary to the predictions of mean field theory, the scaling at small $t$ is sublinear, with exponent $\gamma-1=0.8$. \label{fig:shapefixedS}}
\end{figure}

We have also measured avalanche shapes for avalanches of fixed size $S$. In this case, the expected scaling form is~\cite{Dobrinevski2014, Durin2016}
\begin{equation}
\label{eq:avshapefixedS}
\langle \dot{\epsilon} \rangle_S = \frac{S}{\tau_m} \biggl(\frac{S_m}{S} \biggr)^{1/\gamma} f\biggl(\biggl(\frac{S_m}{S} \biggr)^{1/\gamma}\frac{t}{\tau_m}\biggr)
\end{equation}
where $S_m$ and $\tau_m$ are size and time scales, and $f$ is a universal scaling function. Accordingly, for a given loading condition, mean avalanche shapes at different sizes can be collapsed by rescaling $t$ by $S^{-1/\gamma}$ and $\dot{\epsilon}$ by $S^{1/\gamma-1}$, with the exponent $\gamma=1.8$ as measured from $\langle S\rangle$ {\it vs} $T$ data. We show in Fig.~\ref{fig:shapefixedS}(a) this collapse for pure shear loading in 3D. We have also tested data collapse using the mean field exponent $\gamma=2$ and find a less satisfactory collapse, as shown in Fig.~S3. 
In addition, at early times, the universal shape $f(t)\sim t^{\gamma-1}$ is expected~\cite{Dobrinevski2014}. As shown in Fig.~\ref{fig:shapefixedS}(b), this scaling with $\gamma=1.8$ is valid for all loadings. In contrast, the mean field prediction of linear growth ($\gamma=2$) is clearly in contradiction with the data.

\section*{Conclusion}
We have formulated for the first time a tensorial model of amorphous materials plasticity which captures avalanche dynamics and at the same time reproduces the complex, spatially heterogeneous shear localization patterns which emerge as amorphous materials deform under inhomogeneous loads. Using this model we have demonstrated that avalanches in both 2 and 3 dimensions are characterized by universal, dimension independent critical exponents which are consistently observed across a wide range of loading conditions including heterogeneous and multi-axial loading. The exponents we have obtained, $1/\sigma = 1.95$, $\tau = 1.28$ and $\gamma = 1.8$ differ from the predictions $1/\sigma = 2$, $\tau = 1.5$ and $\gamma = 2$ derived from a mean-field theory of the depinning transition \cite{Dahmen2009} but remain consistent with the depinning paradigm. In fact, our avalanche size distributions are described {\em quantitatively} by corrections to mean field theory~\cite{LeDoussal2012}, with only $\tau$, $\sigma$ and the non-universal critical stress $\Sigma_c$ as fitting parameters. The agreement between our results and the predictions of depinning models outside of the mean field universality class is further seen in the avalanche temporal signals, which are clearly governed by the exponent $\gamma=1.8$ and which have shapes inconsistent with mean field predictions.

Our results provide compelling evidence that the universality class of plastic yield in amorphous materials is not mean field depinning as has been previously claimed, even though the phenomenon may still be considered within the depinning paradigm. We have demonstrated this by using a model that is both truly quasistatic --- a feature which has been shown to be of great importance for measurement of critical exponents~\cite{Liu2016} and which is difficult to ensure in experiments and molecular dynamics simulations --- and also tensorial, thereby being -- unlike scalar models -- directly applicable to real plasticity experiments. The universality of the determined exponents provides an {\em a posteriori} justification of the use of scalar models which are ubiquitous in the statistical physics literature on amorphous plasticity.

A possible explanation for the failure of mean field theory may be related to the fact that the long-range elastic interactions associated with elasto-plasticity are not associated with a positively definite elastic kernel, invalidating a crucial assumption of the renormalization group theory on which  our current understanding of depinning phenomena \cite{LeDoussal2002} is based. Intriguingly, the exponents we measure are close to the values known for depinning of 1D lines with long-range interactions ($\tau=1.25\pm0.05$~\cite{Bonamy2008,Laurson2010}, $1/\sigma=2.1\pm0.08$ \cite{Laurson2010}, $\gamma=1.7$ \cite{Bonamy2008}), and the universal scaling form of \eqref{eq:avalancheDistribution} is a renormalization group prediction valid for long-range 1D depinning~\cite{LeDoussal2012}. This suggests that dimensional reduction via strain localization may be relevant in these systems~\cite{Budrikis2013, Liu2016}. Such strain localization on linear/planar manifolds, which follow directions that are dictated by the macroscopic stress state, may also be an essential factor in the emergence of shear bands which may thus be intimately related to the avalanche dynamics. 


\section*{Acknowledgements}
ZB and SZ are supported by ERC Advanced Grant SIZEFFECTS. SZ acknowledges support from the Academy of Finland FiDiPro progam, project 13282993. MZ acknowledges support from EPSRC under grant EP/J003387/1 and from DFG under grant Za-171/8-1. ZB thanks Ezequiel Ferrero
 and Gianfranco Durin for useful discussions.

\bibliography{plasticity} 

\begin{thebibliography}{51}
\expandafter\ifx\csname natexlab\endcsname\relax\def\natexlab#1{#1}\fi
\expandafter\ifx\csname bibnamefont\endcsname\relax
  \def\bibnamefont#1{#1}\fi
\expandafter\ifx\csname bibfnamefont\endcsname\relax
  \def\bibfnamefont#1{#1}\fi
\expandafter\ifx\csname citenamefont\endcsname\relax
  \def\citenamefont#1{#1}\fi
\expandafter\ifx\csname url\endcsname\relax
  \def\url#1{\texttt{#1}}\fi
\expandafter\ifx\csname urlprefix\endcsname\relax\def\urlprefix{URL }\fi
\providecommand{\bibinfo}[2]{#2}
\providecommand{\eprint}[2][]{\url{#2}}

\bibitem[{\citenamefont{Durin and Zapperi}(2006)}]{Durin2006}
\bibinfo{author}{\bibfnamefont{G.}~\bibnamefont{Durin}} \bibnamefont{and}
  \bibinfo{author}{\bibfnamefont{S.}~\bibnamefont{Zapperi}}, in
  \emph{\bibinfo{booktitle}{The Science of Hysteresis}}, edited by
  \bibinfo{editor}{\bibfnamefont{G.}~\bibnamefont{Bertotti}} \bibnamefont{and}
  \bibinfo{editor}{\bibfnamefont{I.}~\bibnamefont{Mayergoyz}}
  (\bibinfo{publisher}{Elsevier}, \bibinfo{address}{Amsterdam},
  \bibinfo{year}{2006}), vol.~\bibinfo{volume}{2}, pp.
  \bibinfo{pages}{181--267}.

\bibitem[{\citenamefont{Bonamy et~al.}(2008)\citenamefont{Bonamy, Santucci, and
  Ponson}}]{Bonamy2008}
\bibinfo{author}{\bibfnamefont{D.}~\bibnamefont{Bonamy}},
  \bibinfo{author}{\bibfnamefont{S.}~\bibnamefont{Santucci}}, \bibnamefont{and}
  \bibinfo{author}{\bibfnamefont{L.}~\bibnamefont{Ponson}},
  \bibinfo{journal}{Phys. Rev. Lett.} \textbf{\bibinfo{volume}{101}},
  \bibinfo{pages}{045501} (\bibinfo{year}{2008}),
  \urlprefix\url{http://link.aps.org/doi/10.1103/PhysRevLett.101.045501}.

\bibitem[{\citenamefont{Laurson et~al.}(2010)\citenamefont{Laurson, Santucci,
  and Zapperi}}]{Laurson2010}
\bibinfo{author}{\bibfnamefont{L.}~\bibnamefont{Laurson}},
  \bibinfo{author}{\bibfnamefont{S.}~\bibnamefont{Santucci}}, \bibnamefont{and}
  \bibinfo{author}{\bibfnamefont{S.}~\bibnamefont{Zapperi}},
  \bibinfo{journal}{Phys. Rev. E} \textbf{\bibinfo{volume}{81}},
  \bibinfo{pages}{046116} (\bibinfo{year}{2010}),
  \urlprefix\url{http://link.aps.org/doi/10.1103/PhysRevE.81.046116}.

\bibitem[{\citenamefont{Laurson et~al.}(2013)\citenamefont{Laurson, Illa,
  Santucci, Tallakstad, M{\aa}l{\o}y, and Alava}}]{Laurson2013}
\bibinfo{author}{\bibfnamefont{L.}~\bibnamefont{Laurson}},
  \bibinfo{author}{\bibfnamefont{X.}~\bibnamefont{Illa}},
  \bibinfo{author}{\bibfnamefont{S.}~\bibnamefont{Santucci}},
  \bibinfo{author}{\bibfnamefont{K.~T.} \bibnamefont{Tallakstad}},
  \bibinfo{author}{\bibfnamefont{K.~J.} \bibnamefont{M{\aa}l{\o}y}},
  \bibnamefont{and} \bibinfo{author}{\bibfnamefont{M.~J.} \bibnamefont{Alava}},
  \bibinfo{journal}{Nature communications} \textbf{\bibinfo{volume}{4}}
  (\bibinfo{year}{2013}).

\bibitem[{\citenamefont{Argon and Kuo}(1979)}]{Argon1979101}
\bibinfo{author}{\bibfnamefont{A.}~\bibnamefont{Argon}} \bibnamefont{and}
  \bibinfo{author}{\bibfnamefont{H.}~\bibnamefont{Kuo}},
  \bibinfo{journal}{Materials Science and Engineering}
  \textbf{\bibinfo{volume}{39}}, \bibinfo{pages}{101 } (\bibinfo{year}{1979}),
  ISSN \bibinfo{issn}{0025-5416},
  \urlprefix\url{http://www.sciencedirect.com/science/article/pii/0025541679901745}.

\bibitem[{\citenamefont{Argon and Shi}(1983)}]{Argon1983499}
\bibinfo{author}{\bibfnamefont{A.}~\bibnamefont{Argon}} \bibnamefont{and}
  \bibinfo{author}{\bibfnamefont{L.}~\bibnamefont{Shi}}, \bibinfo{journal}{Acta
  Metallurgica} \textbf{\bibinfo{volume}{31}}, \bibinfo{pages}{499 }
  (\bibinfo{year}{1983}), ISSN \bibinfo{issn}{0001-6160},
  \urlprefix\url{http://www.sciencedirect.com/science/article/pii/000161608390038X}.

\bibitem[{\citenamefont{Falk and Langer}(1998)}]{Falk1998}
\bibinfo{author}{\bibfnamefont{M.~L.} \bibnamefont{Falk}} \bibnamefont{and}
  \bibinfo{author}{\bibfnamefont{J.~S.} \bibnamefont{Langer}},
  \bibinfo{journal}{Phys. Rev. E} \textbf{\bibinfo{volume}{57}},
  \bibinfo{pages}{7192} (\bibinfo{year}{1998}),
  \urlprefix\url{http://link.aps.org/doi/10.1103/PhysRevE.57.7192}.

\bibitem[{\citenamefont{Albaret et~al.}(2016)\citenamefont{Albaret, Tanguy,
  Boioli, and Rodney}}]{Albaret2016}
\bibinfo{author}{\bibfnamefont{T.}~\bibnamefont{Albaret}},
  \bibinfo{author}{\bibfnamefont{A.}~\bibnamefont{Tanguy}},
  \bibinfo{author}{\bibfnamefont{F.}~\bibnamefont{Boioli}}, \bibnamefont{and}
  \bibinfo{author}{\bibfnamefont{D.}~\bibnamefont{Rodney}},
  \bibinfo{journal}{Phys. Rev. E} \textbf{\bibinfo{volume}{93}},
  \bibinfo{pages}{053002} (\bibinfo{year}{2016}),
  \urlprefix\url{http://link.aps.org/doi/10.1103/PhysRevE.93.053002}.

\bibitem[{\citenamefont{Baret et~al.}(2002)\citenamefont{Baret, Vandembroucq,
  and Roux}}]{Baret2002}
\bibinfo{author}{\bibfnamefont{J.-C.} \bibnamefont{Baret}},
  \bibinfo{author}{\bibfnamefont{D.}~\bibnamefont{Vandembroucq}},
  \bibnamefont{and} \bibinfo{author}{\bibfnamefont{S.}~\bibnamefont{Roux}},
  \bibinfo{journal}{Phys. Rev. Lett.} \textbf{\bibinfo{volume}{89}},
  \bibinfo{pages}{195506} (\bibinfo{year}{2002}),
  \urlprefix\url{http://link.aps.org/doi/10.1103/PhysRevLett.89.195506}.

\bibitem[{\citenamefont{Picard et~al.}(2002)\citenamefont{Picard, Ajdari,
  Bocquet, and Lequeux}}]{Picard2002}
\bibinfo{author}{\bibfnamefont{G.}~\bibnamefont{Picard}},
  \bibinfo{author}{\bibfnamefont{A.}~\bibnamefont{Ajdari}},
  \bibinfo{author}{\bibfnamefont{L.}~\bibnamefont{Bocquet}}, \bibnamefont{and}
  \bibinfo{author}{\bibfnamefont{F.~m.~c.} \bibnamefont{Lequeux}},
  \bibinfo{journal}{Phys. Rev. E} \textbf{\bibinfo{volume}{66}},
  \bibinfo{pages}{051501} (\bibinfo{year}{2002}),
  \urlprefix\url{http://link.aps.org/doi/10.1103/PhysRevE.66.051501}.

\bibitem[{\citenamefont{Picard et~al.}(2005)\citenamefont{Picard, Ajdari,
  Lequeux, and Bocquet}}]{Picard2005}
\bibinfo{author}{\bibfnamefont{G.}~\bibnamefont{Picard}},
  \bibinfo{author}{\bibfnamefont{A.}~\bibnamefont{Ajdari}},
  \bibinfo{author}{\bibfnamefont{F.~m.~c.} \bibnamefont{Lequeux}},
  \bibnamefont{and} \bibinfo{author}{\bibfnamefont{L.}~\bibnamefont{Bocquet}},
  \bibinfo{journal}{Phys. Rev. E} \textbf{\bibinfo{volume}{71}},
  \bibinfo{pages}{010501} (\bibinfo{year}{2005}),
  \urlprefix\url{http://link.aps.org/doi/10.1103/PhysRevE.71.010501}.

\bibitem[{\citenamefont{Jagla}(2007)}]{Jagla2007}
\bibinfo{author}{\bibfnamefont{E.~A.} \bibnamefont{Jagla}},
  \bibinfo{journal}{Phys. Rev. E} \textbf{\bibinfo{volume}{76}},
  \bibinfo{pages}{046119} (\bibinfo{year}{2007}),
  \urlprefix\url{http://link.aps.org/doi/10.1103/PhysRevE.76.046119}.

\bibitem[{\citenamefont{Shan et~al.}(2008)\citenamefont{Shan, Li, Cheng, Minor,
  Syed~Asif, Warren, and Ma}}]{Shan2008}
\bibinfo{author}{\bibfnamefont{Z.~W.} \bibnamefont{Shan}},
  \bibinfo{author}{\bibfnamefont{J.}~\bibnamefont{Li}},
  \bibinfo{author}{\bibfnamefont{Y.~Q.} \bibnamefont{Cheng}},
  \bibinfo{author}{\bibfnamefont{A.~M.} \bibnamefont{Minor}},
  \bibinfo{author}{\bibfnamefont{S.~A.} \bibnamefont{Syed~Asif}},
  \bibinfo{author}{\bibfnamefont{O.~L.} \bibnamefont{Warren}},
  \bibnamefont{and} \bibinfo{author}{\bibfnamefont{E.}~\bibnamefont{Ma}},
  \bibinfo{journal}{Phys. Rev. B} \textbf{\bibinfo{volume}{77}},
  \bibinfo{pages}{155419} (\bibinfo{year}{2008}),
  \urlprefix\url{http://link.aps.org/doi/10.1103/PhysRevB.77.155419}.

\bibitem[{\citenamefont{Dahmen et~al.}(2009)\citenamefont{Dahmen, Ben-Zion, and
  Uhl}}]{Dahmen2009}
\bibinfo{author}{\bibfnamefont{K.~A.} \bibnamefont{Dahmen}},
  \bibinfo{author}{\bibfnamefont{Y.}~\bibnamefont{Ben-Zion}}, \bibnamefont{and}
  \bibinfo{author}{\bibfnamefont{J.~T.} \bibnamefont{Uhl}},
  \bibinfo{journal}{Phys. Rev. Lett.} \textbf{\bibinfo{volume}{102}},
  \bibinfo{pages}{175501} (\bibinfo{year}{2009}),
  \urlprefix\url{http://link.aps.org/doi/10.1103/PhysRevLett.102.175501}.

\bibitem[{\citenamefont{Wang et~al.}(2009)\citenamefont{Wang, Chan, Xia, Yu,
  Shen, and Wang}}]{Wang2009}
\bibinfo{author}{\bibfnamefont{G.}~\bibnamefont{Wang}},
  \bibinfo{author}{\bibfnamefont{K.}~\bibnamefont{Chan}},
  \bibinfo{author}{\bibfnamefont{L.}~\bibnamefont{Xia}},
  \bibinfo{author}{\bibfnamefont{P.}~\bibnamefont{Yu}},
  \bibinfo{author}{\bibfnamefont{J.}~\bibnamefont{Shen}}, \bibnamefont{and}
  \bibinfo{author}{\bibfnamefont{W.}~\bibnamefont{Wang}},
  \bibinfo{journal}{Acta Materialia} \textbf{\bibinfo{volume}{57}},
  \bibinfo{pages}{6146 } (\bibinfo{year}{2009}), ISSN
  \bibinfo{issn}{1359-6454},
  \urlprefix\url{http://www.sciencedirect.com/science/article/pii/S1359645409005527}.

\bibitem[{\citenamefont{Homer et~al.}(2010)\citenamefont{Homer, Rodney, and
  Schuh}}]{Homer2010}
\bibinfo{author}{\bibfnamefont{E.~R.} \bibnamefont{Homer}},
  \bibinfo{author}{\bibfnamefont{D.}~\bibnamefont{Rodney}}, \bibnamefont{and}
  \bibinfo{author}{\bibfnamefont{C.~A.} \bibnamefont{Schuh}},
  \bibinfo{journal}{Phys. Rev. B} \textbf{\bibinfo{volume}{81}},
  \bibinfo{pages}{064204} (\bibinfo{year}{2010}),
  \urlprefix\url{http://link.aps.org/doi/10.1103/PhysRevB.81.064204}.

\bibitem[{\citenamefont{Sun et~al.}(2010)\citenamefont{Sun, Yu, Jiao, Bai,
  Zhao, and Wang}}]{Sun2010}
\bibinfo{author}{\bibfnamefont{B.~A.} \bibnamefont{Sun}},
  \bibinfo{author}{\bibfnamefont{H.~B.} \bibnamefont{Yu}},
  \bibinfo{author}{\bibfnamefont{W.}~\bibnamefont{Jiao}},
  \bibinfo{author}{\bibfnamefont{H.~Y.} \bibnamefont{Bai}},
  \bibinfo{author}{\bibfnamefont{D.~Q.} \bibnamefont{Zhao}}, \bibnamefont{and}
  \bibinfo{author}{\bibfnamefont{W.~H.} \bibnamefont{Wang}},
  \bibinfo{journal}{Phys. Rev. Lett.} \textbf{\bibinfo{volume}{105}},
  \bibinfo{pages}{035501} (\bibinfo{year}{2010}),
  \urlprefix\url{http://link.aps.org/doi/10.1103/PhysRevLett.105.035501}.

\bibitem[{\citenamefont{Sun and Wang}(2011)}]{Sun2011}
\bibinfo{author}{\bibfnamefont{B.~A.} \bibnamefont{Sun}} \bibnamefont{and}
  \bibinfo{author}{\bibfnamefont{W.~H.} \bibnamefont{Wang}},
  \bibinfo{journal}{Applied Physics Letters} \textbf{\bibinfo{volume}{98}},
  \bibinfo{eid}{201902} (\bibinfo{year}{2011}),
  \urlprefix\url{http://scitation.aip.org/content/aip/journal/apl/98/20/10.1063/1.3592249}.

\bibitem[{\citenamefont{Martens et~al.}(2011)\citenamefont{Martens, Bocquet,
  and Barrat}}]{Martens2011}
\bibinfo{author}{\bibfnamefont{K.}~\bibnamefont{Martens}},
  \bibinfo{author}{\bibfnamefont{L.}~\bibnamefont{Bocquet}}, \bibnamefont{and}
  \bibinfo{author}{\bibfnamefont{J.-L.} \bibnamefont{Barrat}},
  \bibinfo{journal}{Phys. Rev. Lett.} \textbf{\bibinfo{volume}{106}},
  \bibinfo{pages}{156001} (\bibinfo{year}{2011}),
  \urlprefix\url{http://link.aps.org/doi/10.1103/PhysRevLett.106.156001}.

\bibitem[{\citenamefont{Talamali et~al.}(2011)\citenamefont{Talamali,
  Pet\"aj\"a, Vandembroucq, and Roux}}]{Talamali2011}
\bibinfo{author}{\bibfnamefont{M.}~\bibnamefont{Talamali}},
  \bibinfo{author}{\bibfnamefont{V.}~\bibnamefont{Pet\"aj\"a}},
  \bibinfo{author}{\bibfnamefont{D.}~\bibnamefont{Vandembroucq}},
  \bibnamefont{and} \bibinfo{author}{\bibfnamefont{S.}~\bibnamefont{Roux}},
  \bibinfo{journal}{Phys. Rev. E} \textbf{\bibinfo{volume}{84}},
  \bibinfo{pages}{016115} (\bibinfo{year}{2011}),
  \urlprefix\url{http://link.aps.org/doi/10.1103/PhysRevE.84.016115}.

\bibitem[{\citenamefont{Talamali et~al.}(2012)\citenamefont{Talamali, Petäjä,
  Vandembroucq, and Roux}}]{Talamali2012}
\bibinfo{author}{\bibfnamefont{M.}~\bibnamefont{Talamali}},
  \bibinfo{author}{\bibfnamefont{V.}~\bibnamefont{Petäjä}},
  \bibinfo{author}{\bibfnamefont{D.}~\bibnamefont{Vandembroucq}},
  \bibnamefont{and} \bibinfo{author}{\bibfnamefont{S.}~\bibnamefont{Roux}},
  \bibinfo{journal}{Comptes Rendus Mécanique} \textbf{\bibinfo{volume}{340}},
  \bibinfo{pages}{275 } (\bibinfo{year}{2012}), ISSN \bibinfo{issn}{1631-0721},
  \bibinfo{note}{recent Advances in Micromechanics of Materials},
  \urlprefix\url{http://www.sciencedirect.com/science/article/pii/S1631072112000472}.

\bibitem[{\citenamefont{Sun et~al.}(2012)\citenamefont{Sun, Pauly, Tan, Stoica,
  Wang, Kühn, and Eckert}}]{Sun2012}
\bibinfo{author}{\bibfnamefont{B.}~\bibnamefont{Sun}},
  \bibinfo{author}{\bibfnamefont{S.}~\bibnamefont{Pauly}},
  \bibinfo{author}{\bibfnamefont{J.}~\bibnamefont{Tan}},
  \bibinfo{author}{\bibfnamefont{M.}~\bibnamefont{Stoica}},
  \bibinfo{author}{\bibfnamefont{W.}~\bibnamefont{Wang}},
  \bibinfo{author}{\bibfnamefont{U.}~\bibnamefont{Kühn}}, \bibnamefont{and}
  \bibinfo{author}{\bibfnamefont{J.}~\bibnamefont{Eckert}},
  \bibinfo{journal}{Acta Materialia} \textbf{\bibinfo{volume}{60}},
  \bibinfo{pages}{4160 } (\bibinfo{year}{2012}), ISSN
  \bibinfo{issn}{1359-6454},
  \urlprefix\url{http://www.sciencedirect.com/science/article/pii/S1359645412002698}.

\bibitem[{\citenamefont{Budrikis and Zapperi}(2013)}]{Budrikis2013}
\bibinfo{author}{\bibfnamefont{Z.}~\bibnamefont{Budrikis}} \bibnamefont{and}
  \bibinfo{author}{\bibfnamefont{S.}~\bibnamefont{Zapperi}},
  \bibinfo{journal}{Phys. Rev. E} \textbf{\bibinfo{volume}{88}},
  \bibinfo{pages}{062403} (\bibinfo{year}{2013}),
  \urlprefix\url{http://link.aps.org/doi/10.1103/PhysRevE.88.062403}.

\bibitem[{\citenamefont{Salerno and Robbins}(2013)}]{Salerno2013}
\bibinfo{author}{\bibfnamefont{K.~M.} \bibnamefont{Salerno}} \bibnamefont{and}
  \bibinfo{author}{\bibfnamefont{M.~O.} \bibnamefont{Robbins}},
  \bibinfo{journal}{Phys. Rev. E} \textbf{\bibinfo{volume}{88}},
  \bibinfo{pages}{062206} (\bibinfo{year}{2013}),
  \urlprefix\url{http://link.aps.org/doi/10.1103/PhysRevE.88.062206}.

\bibitem[{\citenamefont{Nicolas et~al.}(2014)\citenamefont{Nicolas, Martens,
  Bocquet, and Barrat}}]{Nicolas2014}
\bibinfo{author}{\bibfnamefont{A.}~\bibnamefont{Nicolas}},
  \bibinfo{author}{\bibfnamefont{K.}~\bibnamefont{Martens}},
  \bibinfo{author}{\bibfnamefont{L.}~\bibnamefont{Bocquet}}, \bibnamefont{and}
  \bibinfo{author}{\bibfnamefont{J.-L.} \bibnamefont{Barrat}},
  \bibinfo{journal}{Soft Matter} \textbf{\bibinfo{volume}{10}},
  \bibinfo{pages}{4648} (\bibinfo{year}{2014}),
  \urlprefix\url{http://dx.doi.org/10.1039/C4SM00395K}.

\bibitem[{\citenamefont{Lin et~al.}(2014{\natexlab{a}})\citenamefont{Lin,
  Saade, Lerner, Rosso, and Wyart}}]{Lin2014a}
\bibinfo{author}{\bibfnamefont{J.}~\bibnamefont{Lin}},
  \bibinfo{author}{\bibfnamefont{A.}~\bibnamefont{Saade}},
  \bibinfo{author}{\bibfnamefont{E.}~\bibnamefont{Lerner}},
  \bibinfo{author}{\bibfnamefont{A.}~\bibnamefont{Rosso}}, \bibnamefont{and}
  \bibinfo{author}{\bibfnamefont{M.}~\bibnamefont{Wyart}},
  \bibinfo{journal}{EPL (Europhysics Letters)} \textbf{\bibinfo{volume}{105}},
  \bibinfo{pages}{26003} (\bibinfo{year}{2014}{\natexlab{a}}),
  \urlprefix\url{http://stacks.iop.org/0295-5075/105/i=2/a=26003}.

\bibitem[{\citenamefont{Lin et~al.}(2014{\natexlab{b}})\citenamefont{Lin,
  Lerner, Rosso, and Wyart}}]{Lin2014}
\bibinfo{author}{\bibfnamefont{J.}~\bibnamefont{Lin}},
  \bibinfo{author}{\bibfnamefont{E.}~\bibnamefont{Lerner}},
  \bibinfo{author}{\bibfnamefont{A.}~\bibnamefont{Rosso}}, \bibnamefont{and}
  \bibinfo{author}{\bibfnamefont{M.}~\bibnamefont{Wyart}},
  \bibinfo{journal}{Proceedings of the National Academy of Sciences}
  \textbf{\bibinfo{volume}{111}}, \bibinfo{pages}{14382}
  (\bibinfo{year}{2014}{\natexlab{b}}).

\bibitem[{\citenamefont{Antonaglia
  et~al.}(2014{\natexlab{a}})\citenamefont{Antonaglia, Wright, Gu, Byer,
  Hufnagel, LeBlanc, Uhl, and Dahmen}}]{Antonaglia2014}
\bibinfo{author}{\bibfnamefont{J.}~\bibnamefont{Antonaglia}},
  \bibinfo{author}{\bibfnamefont{W.~J.} \bibnamefont{Wright}},
  \bibinfo{author}{\bibfnamefont{X.}~\bibnamefont{Gu}},
  \bibinfo{author}{\bibfnamefont{R.~R.} \bibnamefont{Byer}},
  \bibinfo{author}{\bibfnamefont{T.~C.} \bibnamefont{Hufnagel}},
  \bibinfo{author}{\bibfnamefont{M.}~\bibnamefont{LeBlanc}},
  \bibinfo{author}{\bibfnamefont{J.~T.} \bibnamefont{Uhl}}, \bibnamefont{and}
  \bibinfo{author}{\bibfnamefont{K.~A.} \bibnamefont{Dahmen}},
  \bibinfo{journal}{Phys. Rev. Lett.} \textbf{\bibinfo{volume}{112}},
  \bibinfo{pages}{155501} (\bibinfo{year}{2014}{\natexlab{a}}),
  \urlprefix\url{http://link.aps.org/doi/10.1103/PhysRevLett.112.155501}.

\bibitem[{\citenamefont{Antonaglia
  et~al.}(2014{\natexlab{b}})\citenamefont{Antonaglia, Xie, Schwarz, Wraith,
  Qiao, Zhang, Liaw, Uhl, and Dahmen}}]{Antonaglia2014a}
\bibinfo{author}{\bibfnamefont{J.}~\bibnamefont{Antonaglia}},
  \bibinfo{author}{\bibfnamefont{X.}~\bibnamefont{Xie}},
  \bibinfo{author}{\bibfnamefont{G.}~\bibnamefont{Schwarz}},
  \bibinfo{author}{\bibfnamefont{M.}~\bibnamefont{Wraith}},
  \bibinfo{author}{\bibfnamefont{J.}~\bibnamefont{Qiao}},
  \bibinfo{author}{\bibfnamefont{Y.}~\bibnamefont{Zhang}},
  \bibinfo{author}{\bibfnamefont{P.~K.} \bibnamefont{Liaw}},
  \bibinfo{author}{\bibfnamefont{J.~T.} \bibnamefont{Uhl}}, \bibnamefont{and}
  \bibinfo{author}{\bibfnamefont{K.~A.} \bibnamefont{Dahmen}},
  \bibinfo{journal}{Scientific Reports} \textbf{\bibinfo{volume}{4}},
  \bibinfo{pages}{4382} (\bibinfo{year}{2014}{\natexlab{b}}).

\bibitem[{\citenamefont{Liu et~al.}(2016)\citenamefont{Liu, Ferrero, Puosi,
  Barrat, and Martens}}]{Liu2016}
\bibinfo{author}{\bibfnamefont{C.}~\bibnamefont{Liu}},
  \bibinfo{author}{\bibfnamefont{E.~E.} \bibnamefont{Ferrero}},
  \bibinfo{author}{\bibfnamefont{F.}~\bibnamefont{Puosi}},
  \bibinfo{author}{\bibfnamefont{J.-L.} \bibnamefont{Barrat}},
  \bibnamefont{and} \bibinfo{author}{\bibfnamefont{K.}~\bibnamefont{Martens}},
  \bibinfo{journal}{Phys. Rev. Lett.} \textbf{\bibinfo{volume}{116}},
  \bibinfo{pages}{065501} (\bibinfo{year}{2016}),
  \urlprefix\url{http://link.aps.org/doi/10.1103/PhysRevLett.116.065501}.

\bibitem[{\citenamefont{Lin et~al.}(2015)\citenamefont{Lin, Gueudre, Rosso, and
  Wyart}}]{Lin2015}
\bibinfo{author}{\bibfnamefont{J.}~\bibnamefont{Lin}},
  \bibinfo{author}{\bibfnamefont{T.}~\bibnamefont{Gueudre}},
  \bibinfo{author}{\bibfnamefont{A.}~\bibnamefont{Rosso}}, \bibnamefont{and}
  \bibinfo{author}{\bibfnamefont{M.}~\bibnamefont{Wyart}},
  \bibinfo{journal}{Phys. Rev. Lett.} \textbf{\bibinfo{volume}{115}},
  \bibinfo{pages}{168001} (\bibinfo{year}{2015}),
  \urlprefix\url{http://link.aps.org/doi/10.1103/PhysRevLett.103.065501}.

\bibitem[{\citenamefont{Sandfeld et~al.}(2015)\citenamefont{Sandfeld, Budrikis,
  Zapperi, and Fernandez~Castellanos}}]{Sandfeld2015}
\bibinfo{author}{\bibfnamefont{S.}~\bibnamefont{Sandfeld}},
  \bibinfo{author}{\bibfnamefont{Z.}~\bibnamefont{Budrikis}},
  \bibinfo{author}{\bibfnamefont{S.}~\bibnamefont{Zapperi}}, \bibnamefont{and}
  \bibinfo{author}{\bibfnamefont{D.}~\bibnamefont{Fernandez~Castellanos}},
  \bibinfo{journal}{Journal of Statistical Mechanics: Theory and Experiment}
  \textbf{\bibinfo{volume}{2015}}, \bibinfo{pages}{P02011}
  (\bibinfo{year}{2015}),
  \urlprefix\url{http://stacks.iop.org/1742-5468/2015/i=2/a=P02011}.

\bibitem[{\citenamefont{Martens et~al.}(2012)\citenamefont{Martens, Bocquet,
  and Barrat}}]{Martens2012}
\bibinfo{author}{\bibfnamefont{K.}~\bibnamefont{Martens}},
  \bibinfo{author}{\bibfnamefont{L.}~\bibnamefont{Bocquet}}, \bibnamefont{and}
  \bibinfo{author}{\bibfnamefont{J.-L.} \bibnamefont{Barrat}},
  \bibinfo{journal}{Soft Matter} \textbf{\bibinfo{volume}{8}},
  \bibinfo{pages}{4197} (\bibinfo{year}{2012}).

\bibitem[{\citenamefont{Zaiser and Moretti}(2005)}]{Zaiser2005}
\bibinfo{author}{\bibfnamefont{M.}~\bibnamefont{Zaiser}} \bibnamefont{and}
  \bibinfo{author}{\bibfnamefont{P.}~\bibnamefont{Moretti}},
  \bibinfo{journal}{Journal of Statistical Mechanics: Theory and Experiment}
  \textbf{\bibinfo{volume}{2005}}, \bibinfo{pages}{P08004}
  (\bibinfo{year}{2005}),
  \urlprefix\url{http://stacks.iop.org/1742-5468/2005/i=08/a=P08004}.

\bibitem[{\citenamefont{Sandfeld and Zaiser}(2014)}]{Sandfeld2014}
\bibinfo{author}{\bibfnamefont{S.}~\bibnamefont{Sandfeld}} \bibnamefont{and}
  \bibinfo{author}{\bibfnamefont{M.}~\bibnamefont{Zaiser}},
  \bibinfo{journal}{Journal of Statistical Mechanics: Theory and Experiment}
  \textbf{\bibinfo{volume}{2014}} (\bibinfo{year}{2014}), ISSN
  \bibinfo{issn}{17425468},
  \urlprefix\url{http://iopscience.iop.org/1742-5468/2014/3/P03014/}.

\bibitem[{\citenamefont{Bulatov and Argon}(1994)}]{Bulatov1994}
\bibinfo{author}{\bibfnamefont{V.~V.} \bibnamefont{Bulatov}} \bibnamefont{and}
  \bibinfo{author}{\bibfnamefont{A.~S.} \bibnamefont{Argon}},
  \bibinfo{journal}{Modelling and Simulation in Materials Science and
  Engineering} \textbf{\bibinfo{volume}{2}}, \bibinfo{pages}{167}
  (\bibinfo{year}{1994}),
  \urlprefix\url{http://stacks.iop.org/0965-0393/2/i=2/a=001}.

\bibitem[{\citenamefont{Anand and Su}(2005)}]{Anand2005}
\bibinfo{author}{\bibfnamefont{L.}~\bibnamefont{Anand}} \bibnamefont{and}
  \bibinfo{author}{\bibfnamefont{C.}~\bibnamefont{Su}},
  \bibinfo{journal}{Journal of the Mechanics and Physics of Solids}
  \textbf{\bibinfo{volume}{53}}, \bibinfo{pages}{1362} (\bibinfo{year}{2005}).

\bibitem[{\citenamefont{Su and Anand}(2006)}]{Su2006}
\bibinfo{author}{\bibfnamefont{C.}~\bibnamefont{Su}} \bibnamefont{and}
  \bibinfo{author}{\bibfnamefont{L.}~\bibnamefont{Anand}},
  \bibinfo{journal}{Acta Materialia} \textbf{\bibinfo{volume}{54}},
  \bibinfo{pages}{179} (\bibinfo{year}{2006}).

\bibitem[{\citenamefont{Yang et~al.}(2006)\citenamefont{Yang, Mota, and
  Ortiz}}]{Yang2006}
\bibinfo{author}{\bibfnamefont{Q.}~\bibnamefont{Yang}},
  \bibinfo{author}{\bibfnamefont{A.}~\bibnamefont{Mota}}, \bibnamefont{and}
  \bibinfo{author}{\bibfnamefont{M.}~\bibnamefont{Ortiz}},
  \bibinfo{journal}{Computational Mechanics} \textbf{\bibinfo{volume}{37}},
  \bibinfo{pages}{194} (\bibinfo{year}{2006}).

\bibitem[{\citenamefont{Homer and Schuh}(2009)}]{Homer2009}
\bibinfo{author}{\bibfnamefont{E.~R.} \bibnamefont{Homer}} \bibnamefont{and}
  \bibinfo{author}{\bibfnamefont{C.~A.} \bibnamefont{Schuh}},
  \bibinfo{journal}{Acta Materialia} \textbf{\bibinfo{volume}{57}},
  \bibinfo{pages}{2823 } (\bibinfo{year}{2009}), ISSN
  \bibinfo{issn}{1359-6454},
  \urlprefix\url{http://www.sciencedirect.com/science/article/pii/S1359645409001311}.

\bibitem[{\citenamefont{Schuh and Lund}(2003)}]{Schuh2003}
\bibinfo{author}{\bibfnamefont{C.}~\bibnamefont{Schuh}} \bibnamefont{and}
  \bibinfo{author}{\bibfnamefont{A.}~\bibnamefont{Lund}},
  \bibinfo{journal}{Nature materials} \textbf{\bibinfo{volume}{2}},
  \bibinfo{pages}{449} (\bibinfo{year}{2003}).

\bibitem[{\citenamefont{D.}(1957)}]{Eshelby1957}
\bibinfo{author}{\bibfnamefont{E.~J.} \bibnamefont{D.}},
  \bibinfo{journal}{Proceedings of the Royal Society of London. Series A,
  Mathematical and Physical Sciences} \textbf{\bibinfo{volume}{241}},
  \bibinfo{pages}{pp. 376} (\bibinfo{year}{1957}), ISSN
  \bibinfo{issn}{00804630}, \urlprefix\url{http://www.jstor.org/stable/100095}.

\bibitem[{\citenamefont{Vaidyanathan et~al.}(2001)\citenamefont{Vaidyanathan,
  Dao, Ravichandran, and Suresh}}]{Vaidyanathan2001}
\bibinfo{author}{\bibfnamefont{R.}~\bibnamefont{Vaidyanathan}},
  \bibinfo{author}{\bibfnamefont{M.}~\bibnamefont{Dao}},
  \bibinfo{author}{\bibfnamefont{G.}~\bibnamefont{Ravichandran}},
  \bibnamefont{and} \bibinfo{author}{\bibfnamefont{S.}~\bibnamefont{Suresh}},
  \bibinfo{journal}{Acta Materialia} \textbf{\bibinfo{volume}{49}},
  \bibinfo{pages}{3781 } (\bibinfo{year}{2001}), ISSN
  \bibinfo{issn}{1359-6454},
  \urlprefix\url{http://www.sciencedirect.com/science/article/pii/S1359645401002634}.

\bibitem[{\citenamefont{Ramamurty et~al.}(2005)\citenamefont{Ramamurty, Jana,
  Kawamura, and Chattopadhyay}}]{Ramamurty2005}
\bibinfo{author}{\bibfnamefont{U.}~\bibnamefont{Ramamurty}},
  \bibinfo{author}{\bibfnamefont{S.}~\bibnamefont{Jana}},
  \bibinfo{author}{\bibfnamefont{Y.}~\bibnamefont{Kawamura}}, \bibnamefont{and}
  \bibinfo{author}{\bibfnamefont{K.}~\bibnamefont{Chattopadhyay}},
  \bibinfo{journal}{Acta Materialia} \textbf{\bibinfo{volume}{53}},
  \bibinfo{pages}{705 } (\bibinfo{year}{2005}), ISSN \bibinfo{issn}{1359-6454},
  \urlprefix\url{http://www.sciencedirect.com/science/article/pii/S1359645404006329}.

\bibitem[{\citenamefont{Shi and Falk}(2007)}]{Shi2007}
\bibinfo{author}{\bibfnamefont{Y.}~\bibnamefont{Shi}} \bibnamefont{and}
  \bibinfo{author}{\bibfnamefont{M.~L.} \bibnamefont{Falk}},
  \bibinfo{journal}{Acta Materialia} \textbf{\bibinfo{volume}{55}},
  \bibinfo{pages}{4317 } (\bibinfo{year}{2007}), ISSN
  \bibinfo{issn}{1359-6454},
  \urlprefix\url{http://www.sciencedirect.com/science/article/pii/S135964540700239X}.

\bibitem[{\citenamefont{Maloney and Robbins}(2009)}]{Maloney2009}
\bibinfo{author}{\bibfnamefont{C.~E.} \bibnamefont{Maloney}} \bibnamefont{and}
  \bibinfo{author}{\bibfnamefont{M.~O.} \bibnamefont{Robbins}},
  \bibinfo{journal}{Phys. Rev. Lett.} \textbf{\bibinfo{volume}{102}},
  \bibinfo{pages}{225502} (\bibinfo{year}{2009}),
  \urlprefix\url{http://link.aps.org/doi/10.1103/PhysRevLett.102.225502}.

\bibitem[{\citenamefont{Chen and Lin}(2010)}]{Chen2010}
\bibinfo{author}{\bibfnamefont{K.}~\bibnamefont{Chen}} \bibnamefont{and}
  \bibinfo{author}{\bibfnamefont{J.}~\bibnamefont{Lin}}, \bibinfo{journal}{Int.
  J. Plasticity} \textbf{\bibinfo{volume}{26}}, \bibinfo{pages}{1645}
  (\bibinfo{year}{2010}).

\bibitem[{\citenamefont{Le~Doussal and Wiese}(2012)}]{LeDoussal2012}
\bibinfo{author}{\bibfnamefont{P.}~\bibnamefont{Le~Doussal}} \bibnamefont{and}
  \bibinfo{author}{\bibfnamefont{K.~J.} \bibnamefont{Wiese}},
  \bibinfo{journal}{Phys. Rev. E} \textbf{\bibinfo{volume}{85}},
  \bibinfo{pages}{061102} (\bibinfo{year}{2012}),
  \urlprefix\url{http://link.aps.org/doi/10.1103/PhysRevE.85.061102}.

\bibitem[{\citenamefont{Dobrinevski et~al.}(2014)\citenamefont{Dobrinevski,
  Doussal, and Wiese}}]{Dobrinevski2014}
\bibinfo{author}{\bibfnamefont{A.}~\bibnamefont{Dobrinevski}},
  \bibinfo{author}{\bibfnamefont{P.~L.} \bibnamefont{Doussal}},
  \bibnamefont{and} \bibinfo{author}{\bibfnamefont{K.~J.} \bibnamefont{Wiese}},
  \bibinfo{journal}{EPL (Europhysics Letters)} \textbf{\bibinfo{volume}{108}},
  \bibinfo{pages}{66002} (\bibinfo{year}{2014}),
  \urlprefix\url{http://stacks.iop.org/0295-5075/108/i=6/a=66002}.

\bibitem[{\citenamefont{Durin et~al.}(2016)\citenamefont{Durin, Bohn, Correa,
  Sommer, Doussal, and Wiese}}]{Durin2016}
\bibinfo{author}{\bibfnamefont{G.}~\bibnamefont{Durin}},
  \bibinfo{author}{\bibfnamefont{F.}~\bibnamefont{Bohn}},
  \bibinfo{author}{\bibfnamefont{M.}~\bibnamefont{Correa}},
  \bibinfo{author}{\bibfnamefont{R.}~\bibnamefont{Sommer}},
  \bibinfo{author}{\bibfnamefont{P.~L.} \bibnamefont{Doussal}},
  \bibnamefont{and} \bibinfo{author}{\bibfnamefont{K.}~\bibnamefont{Wiese}},
  \bibinfo{journal}{arXiv preprint arXiv:1601.01331}  (\bibinfo{year}{2016}).

\bibitem[{\citenamefont{Le~Doussal et~al.}(2002)\citenamefont{Le~Doussal,
  Wiese, and Chauve}}]{LeDoussal2002}
\bibinfo{author}{\bibfnamefont{P.}~\bibnamefont{Le~Doussal}},
  \bibinfo{author}{\bibfnamefont{K.~J.} \bibnamefont{Wiese}}, \bibnamefont{and}
  \bibinfo{author}{\bibfnamefont{P.}~\bibnamefont{Chauve}},
  \bibinfo{journal}{Phys. Rev. B} \textbf{\bibinfo{volume}{66}},
  \bibinfo{pages}{174201} (\bibinfo{year}{2002}),
  \urlprefix\url{http://link.aps.org/doi/10.1103/PhysRevB.66.174201}.

\end{thebibliography}




\end{document}